\documentclass{llncs}
\usepackage[utf8]{inputenc}
\usepackage[T1]{fontenc}
\usepackage[hyphens]{url}
\usepackage{hyperref}
\usepackage{amsmath,amssymb}

\newcommand{\ChapSevenAuthorOne}{{Emmanuel Thomé}}
\newcommand{\ChapSevenAuthorOneAffiliation}{Inria, Nancy, France}

\newcommand{\F}{\mathbb{F}}

\DeclareMathOperator{\rank}{rank}
\DeclareMathOperator{\Expect}{E}

\begin{document}

\title{A modified block Lanczos algorithm with fewer vectors}

\makeatletter
\def\thefootnote{\@fnsymbol\c@footnote}
\stepcounter{footnote}
    \footnotetext{This article is based on 
        work by the author contributed as chapter~7 in 
        \emph{Topics in
            Computational
        Number Theory inspired by Peter L. Montgomery}, by
        Joppe W. Bos and Arjen K. Lenstra, to be published by Cambdridge
    University Press.}%
\stepcounter{footnote}%
\footnotetext{\today; Version for this file: \texttt{%
        d76dfc6

}}
\makeatother

\author{Emmanuel Thomé}
\institute{%
INRIA Nancy / LORIA\\
615 rue du jardin botanique\\
54600 Villers-lès-Nancy}
\maketitle

\begin{abstract}

\nocite{BoLe16}

The block Lanczos algorithm proposed by Peter Montgomery is an efficient
means to tackle the sparse linear algebra problem which arises in the
context of the number field sieve factoring algorithm and its
predecessors. We present here a modified version of the algorithm, which
incorporates several improvements: we discuss how to efficiently handle
homogeneous systems and how to reduce the number of vectors stored in the
course of the computation. We also provide heuristic justification for
the success probability of our modified algorithm.

While the overall complexity and expected number of steps of the block
Lanczos is not changed by the modifications presented in this article, we
expect these to be useful for implementations of the block Lanczos
algorithm where the storage of auxiliary vectors sometimes has a
non-negligible cost.
\end{abstract}

\let\thechapter\undefined
\def\processingarticle{x}
\def\bfv{\mathbf{v}}
\def\bfc{\mathbf{c}}
\def\bfp{\mathbf{p}}
\def\bfw{\mathbf{w}}
\def\bft{\mathbf{t}}
\def\bfd{\mathbf{d}}
\def\bfX{\mathbf{X}}
\def\bfS{\mathbf{S}}
\def\bfx{\mathbf{x}}
\def\bfSigma{\mathbf{\Sigma}}
\def\bfdelta{\mathbf{\delta}}
\def\bfDelta{\mathbf{\Delta}}
\def\cO{\mathop{\mathcal{O}}}
\def\barmatrix#1#2#3#4{\left(\begin{array}{c|c}#1&#2\\\hline#3&#4\end{array}\right)}
\def\barvector#1#2{\left(\begin{array}{c|c}#1&#2\end{array}\right)}

\ifx\thechapter\undefined\else

\author{\ChapSevenAuthorOne}

\chapterauthor{\ChapSevenAuthorOne}

\setcounter{chapter}{6}

\chapter{The block Lanczos algorithm}\label{chap7:blocklanczos}

\contributor{\ChapSevenAuthorOne
\affiliation{\ChapSevenAuthorOneAffiliation}}

We present the block Lanczos algorithm proposed by Peter Montgomery,
which is an efficient means to tackle the sparse linear algebra problem which
arises in the context of the number field sieve factoring algorithm and
its predecessors. The presentation incorporates some
simplifications and improvements.

\fi

\section{Linear systems for integer factoring}
\label{sec:blocklanczos:intro}

For factoring a composite integer $N$, algorithms based on the technique
of combination of congruences look for several pairs of integers $(x,y)$
such that $$x^2\equiv y^2\mod N.$$ This equality is hoped to be non
trivial for at least one of the obtained pairs, letting $\gcd(x-y,N)$
unveil a factor of the integer $N$.

Several algorithms use this strategy: the CFRAC algorithm,  the
quadratic sieve and its variants, and the number field sieve. Pairs
$(x,y)$ as above are obtained by combining relations which have been collected
as a step of these algorithms. Relations are written multiplicatively as
a set of valuations. All the algorithms considered seek a multiplicative
combination of these relations which can be rewritten as an equality of
squares. This is achieved by solving a system of linear equations defined
over $\F_2$, where equations are parity constraints on each valuation
considered, and unknowns indicate whether or not relations are to be
selected as part of the combination.

We are therefore facing a linear algebra problem. Writing the relations
collected as the rows of a matrix $M$ with coefficients in $\F_2$, we are
to find several solutions to the homogeneous linear system $$x^TM=0.$$ To
fix notations, we let the matrix $M$ be square of size $N\times N$. It is
noteworthy that the matrix $M$ is extremely sparse, as can be illustrated
by data from some factoring experiments: for the factoring of RSA-512 in
1999, the matrix $M$ had $N\approx 7\times10^6$ and $62$ non-zero
coefficients per row, and for the RSA-768 factorization in 2009, the
matrix $M$ had $N\approx 2\times10^8$ and $144$ non-zero coefficients per
row.

This sparsity property can be exploited to yield efficient algorithms
which solve the linear system in a ``black-box'' fashion, that is,
without ever modifying the matrix $M$. The only access to the matrix $M$
which is allowed to such algorithms is the operation of multiplying $M$
(or its transpose) by a vector, and obtain the result. The interesting
black-box algorithms are those which solve the linear system using at
most $O(N)$ times this operation. For sparse matrices, this approach is
considerably cheaper than ``dense'' algorithms which do not exploit the
sparsity property, with regard to both the time and space complexity
(which would, for dense algorithms, be $O(N^\omega)$ and $O(N^2)$).

\section{The standard Lanczos algorithm}
\label{sec:blocklanczos:lanczos}

Dealing with sparse linear systems is an important topic which goes
beyond computational number theory.  Among the sparse algorithms which
can be employed (reviewed in early works such as~\cite{C:LaMOdl90a}), we
find the conjugate gradient and the Lanczos algorithms, which were both
originally stated in the context of solving numerical systems occurring,
for example, in the context of the solution of partial differential
equations. With some adaptation work, it is possible to use these
algorithms over finite fields, with limitations which we will
mention in §\ref{sec:blocklanczos:char2}.
The Wiedemann algorithm~\cite{Wiedemann86} was proposed as a
method particularly well adapted to finite fields. We will discuss in
§\ref{sec:blocklanczos:recent} how it
compares with the Lanczos and block Lanczos algorithms.

As a first step towards presenting the block Lanczos algorithm, we give
here an overview of how the standard Lanczos algorithm can be used to
solve homogeneous or inhomogeneous linear systems over finite fields.
Arguments which appear in the justification of the standard Lanczos are
also important to the block Lanczos context, which explains this
preliminary overview.  Within this section, we assume that the base field
is $\F_p$ for some prime $p$.

Briefly put, the Lanczos algorithm is the Gram-Schmidt orthogonalization
process applied to a Krylov subspace. We need to work with a symmetric
matrix $A$ defined over $\F_p$.  Different problems can be stated, for
example depending on whether we intend to solve a homogeneous or
inhomogeneous linear system. Another distinction comes from the linear
system which we want to solve in the first place. While in some cases it
does indeed define a symmetric matrix $A$, it may also be that we form
$A$ as $A=MM^T$, and solve a linear system involving $A$ as a derived
means of solving one involving $M$. Such a strategy would be natural in
the prospect of solving the linear systems as defined in
§\ref{sec:blocklanczos:intro}. In that case, the matrix $A$ is never actually
computed, and the black box ``multiplication by $A$'' is instead realized
as the composition
of the two black boxes multiplying by $M^T$ and $M$.

For expository purposes, we assume in
this section that we have a right-hand side vector $b\in\F_p^N$,
and intend to solve for $x\in\F_p^{N}$ the equation
$$Ax=b.$$

The matrix $A$ being symmetric, we may consider
the inner product defined from $A$ as $v^TAw$ for 
vectors $v,w\in\F_p^N$. We say that $v$ and $w$ are $A$-orthogonal
whenever $v^TAw=0$. A vector is $A$-isotropic if it is $A$-orthogonal to
itself.

The Lanczos algorithm focuses on the sequence of Krylov subspaces of $\F_p^N$
defined as $V_i=\left\langle v_0,Av_0,A^2v_0,\ldots,A^{i}v_0\right\rangle$,
where $v_0=b$.
It is clear that the sequence of subspaces
$(V_i)_{i\geq0}$ is
strictly increasing up to some index, and then stationary.

We define a sequence of vectors $(v_i)_{i\geq0}$, 
computed so as to satisfy the two following conditions:
\begin{align}
    \label{eq:lanczos:condition-orth} v_i &\mathrel{\text{is $A$-orthogonal
to}}  v_j\ \text{whenever}\ i\not=j,\\
\label{eq:lanczos:condition-span} V_i &=\left\langle v_0,\ldots,v_{i}\right\rangle.
\end{align}
We proceed by induction, and assume that a sequence of vectors $v_0$ to
$v_i$ has been computed so that the two conditions above hold.
We now see how to compute $v_{i+1}$.
We begin by noting (using condition~\eqref{eq:lanczos:condition-span}
inductively) that $$V_{i+1}= \langle v_0\rangle+AV_i=\langle
v_0\rangle +AV_{i-1}+\langle Av_i\rangle=V_i+\langle Av_i\rangle$$
so
that setting $v_{i+1}$ to be any vector within the affine subspace
$V_i+Av_i$ fulfils condition~\eqref{eq:lanczos:condition-span} for
index $i+1$. In
order to satisfy condition~\eqref{eq:lanczos:condition-orth}, we let:
$$
    v_{i+1}
        = Av_i - \sum_{j\leq i}\frac{v_j^TA^2v_i}{v_j^TAv_j}v_j.
            $$
We leave for further discussion the important question of the
non-degeneracy of the denominators in the expression of $v_{i+1}$.

It turns out that the equation above defining $v_{i+1}$ can be
simplified.
Indeed, because $Av_j\in V_{j+1}$, we have that
$v_j^TA^2v_i=0$ whenever $j<i-1$. This implies that only two terms in
the sum above are non-zero, yielding the following shorter equation for
defining $v_{i+1}$:
\begin{gather*}
    v_{i+1} = Av_i - c_{i+1,i} v_i - c_{i+1,i-1} v_{i-1},\\
    c_{i+1,i} =
    \frac{v_i^TA^2v_i}{v_i^TAv_i}v_i,\quad
    c_{i+1,i-1} = 
    \frac{v_{i-1}^TA^2v_{i}}{v_{i-1}^TAv_{i-1}}v_{i-1}.
\end{gather*}
Note also that we have $Av_{i-1} \in v_i + V_{i-1}$,
so that $v_{i-1}^TA^2v_{i}=v_i^TAv_i$. We can
then simplify the expression of $c_{i+1,i-1}$ as
$$c_{i+1,i-1}
=\frac{v_i^TAv_{i}}{v_{i-1}^TAv_{i-1}}v_{i-1}.$$

The sequence of vectors $(v_i)_{i\geq0}$ can thus be computed
with a simple recurrence procedure, requiring only a short amount of
history to be updated from each iteration to the next (namely, the
vectors $v_{i+1}$ and $v_{i}$ as well as the scalar $v_{i}^TAv_{i}$). 

We now discuss the termination of the computation of the sequence of
vectors $(v_i)_{i\geq0}$. It is clear that $v_{i+1}$ can be computed only
as long as the following condition holds:
\begin{equation}
    \label{eq:lanczos:condition-non-isotropic}
    \forall j\leq i,\ v_j^TAv_j\not=0.
\end{equation}
We assume
that condition~\eqref{eq:lanczos:condition-non-isotropic} holds until
some index $m$ (not included), and that $v_{m}=0$. This implies
$V_{m}=V_{m-1}$. Define now $x$ as
$$x=\sum_{i<m}\frac{v_i^Tb}{v_i^TAv_i}v_i.$$
By construction\footnote{This argument uses the fact that we have chosen
$v_0=b$. Had we chosen $v_0$ arbitrarily, then we would need to assume
$b\in V_m$.}, we have $Ax-b\in V_{m-1}$. Vectors $v_i$ for indices $i<m$
form an $A$-orthogonal basis of $V_{m-1}$, therefore $(Ax-b)^TAv_i=0$ for all
$i<m$ because of the expression of $x$. It follows from~\eqref{eq:lanczos:condition-orth}
and~\eqref{eq:lanczos:condition-non-isotropic} that we have $Ax=b$.
Computing the summands of $x$ can be done at the same
time as the sequence $(v_i)_{i\geq0}$ is computed, adding the need for
one extra vector of $\F_p^N$.

Condition~\eqref{eq:lanczos:condition-non-isotropic} may fail to hold
without reaching  $v_{m}=0$, however. This is because the positive characteristic
setting does not forbid $A$-isotropic vectors: it may happen that
$v_m^TAv_m=0$ without $v_{m}=0$. In this case, the algorithm fails. In~\cite{EbKa97},
Eberly and Kaltofen show that
condition~\eqref{eq:lanczos:condition-non-isotropic} is equivalent to the
matrix $H(A,b)=(b^TA^{i+j+1}b)_{0\leq i<m}$ being of generic rank
profile (all leading principal minors are non-zero). They further show
how it is possible to control the failure probability with appropriate
randomization, under the assumption that the coefficient field is large
enough.

\section{The case of characteristic two}
\label{sec:blocklanczos:char2}
When $p=2$, the standard Lanczos algorithm cannot work, as $A$-isotropic
vectors are bound to occur.  This problem is an incurable failure
condition in the finite field case, but an analogous mishap can also be
encountered in the numerical case: if some $v_i^TAv_i$ happens to be very
close to zero, then numerical instability occurs.

Techniques to address this issue have been proposed in the numerical
context quite early on, namely the look-ahead Lanczos
algorithm~\cite{PaTaLi85} which suggests to compute $v_{i+1}$ from
several of the previous iterates.
In the context of integer factorization and
linear systems defined over $\F_2$, early techniques suggested, e.g.
in~\cite{C:LaMOdl90a}, to overcome the issue of $A$-isotropic vectors
were quite inefficient, requiring for example to do all computations in
a field $\F_{2^k}$ for some $k$.
Coppersmith~\cite{Coppersmith93a} and Montgomery~\cite{EC:Montgomery95},
in a somewhat simpler form, proposed to efficiently solve this problem by
taking inspiration from the look-ahead technique, and more importantly by
considering several vectors simultaneously.

The theoretical benefit is that if
we consider a block of $n$ vectors $\bfv $ (represented by a
matrix of size $N\times n$), the matrix $\bfv ^TA\bfv $ might
fail to be invertible, but its rank defect may be expected to be
reasonably small, thereby allowing the algorithm to proceed.

Considering blocks of vectors is also a great practical benefit when
dealing with sparse matrices defined over $\F_2$. Multiplying a sparse
matrix by a vector requires, for each matrix coefficient, to access a
single coefficient (hence a single bit) of the input vector. Despite the
fact that memory access probably reaches the nearby bits of the input
vector as well, these do not matter and one expects that their value is
most often discarded. When a block of $n$ vectors is considered, and $n$
is equal to the machine word size (say, $n=64$), then there is a
natural alternative way to proceed. Storing blocks of vectors as
$N$-element arrays of $n$-bit machine words, it is possible to compute
simultaneously the product of a sparse matrix by a block of vectors in
essentially the same number of distinct memory accesses than required for
doing a single matrix-times-vector operation. The question is then
whether such an approach leads to a modification of the Lanczos algorithm
which requires fewer iterations.

\section{Orthogonalizing a sequence of subspaces}

The key to the block Lanczos algorithm is the idea of considering a
sequence of subspaces of dimension larger than 1. We use boldface letters
to denote blocks of $n$ vectors, and the notation
$\langle\bfv \rangle$ denotes the subspace of $\F_p^N$ spanned by
the $n$ columns of $\bfv$. We extend this trivially to
$\langle\bfv_0,\bfv_1\rangle$
As in the case of the standard Lanczos,
we define notions which are related to the inner product defined by the
matrix $A$. We say that spaces $\langle\bfv \rangle$ and
$\langle\bfw \rangle$ are $A$-orthogonal whenever $\bfv ^TA\bfw =0$. It is
clear that $\bfv ^TA\bfw $ is an $n\times n$ matrix with
coefficients in $\F_p$.

We first describe a naive extension of the Lanczos algorithm to the
setting of blocks of vectors, and explain why it does not work (at least
not if $p$ may be small).
We need to define an analogue to the sequence of mutually orthogonal
vectors $(v_i)_{i\geq0}$ considered in the standard Lanczos algorithm.
Let us fix an arbitrary vector block  $\bfv_0$ as a starting point (to be
discussed in §\ref{sec:blocklanczos:termination}).
We may attempt to define a sequence of vector spaces with
$\bfv_i^TA\bfv_i$ non-singular as follows.
\begin{itemize}
    \item Set $\displaystyle\bft =
        A\bfv_i-
        \sum_{j\leq i}
        \mathbf v_j
        (\mathbf v_j^TA\mathbf v_j)^{-1}
        (\mathbf v_j^TA^2\mathbf v_i)
        $.
    \item Define $\bfv_{i+1}$ as a maximal set of columns within
        $\bft $ so that $\bfv_{i+1}^TA\bfv_{i+1}$ is
        invertible.
\end{itemize}
The key problem with the approach above is that $\bfv_{i+1}$ is
a block of possibly fewer vectors than $\bfv_{i}$: when
$\bft ^TA\bft $ above is not of full rank, some vectors are
discarded and not selected in $\bfv_{i+1}$. This implies that
after
some steps, the expected dimension of the block
$\langle\bfv_i\rangle$ collapses to zero, with no further progress
possible. (A rule of thumb expecting a rank defect of 1 with
probability $\frac1p$ predicts that no more than $np$ steps can be done
before this collapse.)

To address this issue, Montgomery suggested an idea related to 
the look-ahead Lanczos~\cite{PaTaLi85} (but apparently discovered
independently): allow to build orthogonal subspaces
from a larger number of the previous iterates.
For notational
ease, we depart slightly here from the notations used
in~\cite{EC:Montgomery95}.
We define sequences $(\bfv_i)_{i\geq0}$, $(\bfd_i)_{i\geq0}$, and
$(\bfw_i)_{i\geq0}$, where $\bfv_i\in\F_p^{N\times n}$,
$\bfw_i\in\F_p^{N\times n}$, and $\bfd_i\in\F_p^{n\times n}$
diagonal with entries in $\{0,1\}$.
We require, for all $i\geq0$:
\begin{align}
    \notag \bfw_i&=\bfv_i\bfd_i,\\
\label{eq:blocklanczos:condition-minor}
[\bfw_i^TA\bfw_i]_{\bfd_i}&\not=0\text{ (principal minor
marked by $\bfd_i$)}\\
\label{eq:blocklanczos:condition-orth}
\bfw_j^TA\bfv_i&=0\text{ whenever $j<i$}.
\end{align}
The diagonal matrix
$\bfd_i$ essentially encodes the choice of a subset of
$\{1,\ldots,n\}$, which justifies the notation
$[\bfw_i^TA\bfw_i]_{\bfd_i}$ for the principal minor
attached to this set
(note that we have $[\bfv_i^TA\bfv_i]_{\bfd_i}=
[\bfw_i^TA\bfw_i]_{\bfd_i}$).
It is clear that
condition~\eqref{eq:blocklanczos:condition-orth} also implies
$\bfw_i^TA\bfw_j=0$ whenever $i\not=j$.

In Montgomery's algorithm, the sequence of orthogonal subspaces is the
sequence $\bfw_i$, which are formed from as many columns from $\bfv_i$ as
possible (condition~\ref{eq:blocklanczos:condition-minor} imposes that
the inner product defined by $A$ is non-degenerate on
$\langle\bfw_i\rangle$). Vectors from $\bfv_i$ which are \emph{not}
selected in $\bfw_i$, instead of being dropped, are considered again
for selection in the next iterations.

As for the standard Lanczos algorithm, we explain how the conditions
above can be satisfied with an explicit inductive construction. The
starting point of each iteration is the vector block $\bfv_i$. The first
step is to compute  $\bfd_i$ (and hence $\bfw_i$) so as to satisfy
condition~\eqref{eq:blocklanczos:condition-minor}.
In a second step, we
compute $\bfv_{i+1}$ so as to satisfy
condition~\eqref{eq:blocklanczos:condition-orth}.

\section{Construction of the next iterate}
We first discuss how to compute $\bfd_i$ (and hence
$\bfw_i$) from $\bfv_i$. 
    We need the following lemma:
\begin{lemma}
    \label{lem:rank-sym}
    Let $\bfX \in\F_p^{n\times n}$ be a symmetric matrix of rank
    $r$, and $S\subset\{1,\ldots,n\}$ be indices of $r$ independent
    columns of $\bfX $. Then the principal minor $[\bfX ]_S$ is
    non-zero.
\end{lemma}
To see this, assume without loss of generality that $S=\{1,\ldots,r\}$.
Columns of indices $r+1$ and above can be expressed as combinations of
the first $r$ columns. 
We may write a matrix $\bfSigma=\barmatrix{1_r}{\ast}{0}{1_{n-r}}$ so
that $\bfX \bfSigma$ has only its $r$ first columns non zero.
The matrix $\mathbf X'=\bfSigma^T\bfX \bfSigma$ has its last $n-r$ rows
and columns equal to zero,
so that only its
leading $r\times r$ submatrix is non-zero. Since
$\bfX '$ has rank $r$ and this submatrix coincides with the leading
$r\times r$ submatrix of $\bfX $, this is saying that
$[\bfX ]_S\not=0$, as claimed.
\medskip

Lemma~\ref{lem:rank-sym} implies that computing $\bfd_i$ so as to
satisfy condition~\eqref{eq:blocklanczos:condition-minor}
only amounts to Gaussian elimination on the $n\times n$ matrix
$\bfv_i^TA\bfv_i$.

An inverse of the submatrix whose row and column indices are encoded by
$\bfd_i$ can be computed from the same Gaussian elimination procedure.
Therefore, we assume that a by-product of the computation of
$\bfd_i$ is an $n\times n$ matrix $\bfw_i^{\text{inv}}$ such that:
\begin{align*}
    \bfw_i^{\text{inv}} &= \bfw_i^{\text{inv}}\bfd_i = \bfd_i\bfw_i^{\text{inv}},\\
    \bfd_i &=
    \bfw_i^{\text{inv}}(\bfw_i^TA\bfw_i)=
    \bfw_i^{\text{inv}}(\bfv_i^TA\bfv_i)\bfd_i
\end{align*}
The former condition above expresses the fact that $\bfw_i^{\text{inv}}$ is zero
outside the row and column indices encoded by $\bfd_i$, while the latter
expresses the fact that it is an inverse to the corresponding submatrix.
Note that this construction implies that $\bfw_i^{\text{inv}}$ is symmetric.
\medskip

Assuming $\bfd_i$ and $\bfw_i$ have been derived from
$\bfv_i$, we now build $\bfv_{i+1}$ from $A\bfw_i$ and $\bfv_i$. Given that
$A\bfw_i$ has, by construction, $n-\rank(\bfd_i)$ zero columns, we
complete it with the columns of $\bfv_i$ which were not selected in
$\bfw_i$. We then write $\bfv_{i+1}$
as follows.
\begin{align*}
    \bft &= A\bfv_i\bfd_i + \bfv_i(1-\bfd_i),\\
    \bfv_{i+1} &= \bft - \sum_{j\leq i} \bfw_j\bfw_j^{\text{inv}}\bfw_j^TA\bft.
\end{align*}
It is clear from the quantities computed so far that $\bfv_{i+1}$ is
$A$-orthogonal to $\bfw_j$ for all $j\leq i$, which is
condition~\eqref{eq:blocklanczos:condition-orth}.  We remark that we
could have used, as Montgomery does, the vector block
$A\bfw_i+\bfv_i=\bft+\bfw_i$ instead of the value chosen above for
$\bft$, and this would have led to the same value for
$\bfv_{i+1}$.

\section{Simplifying the recurrence equation}

The previous section defines a complete set of equations for determining
$\bfv_i$.  However the expression above for $\bfv_{i+1}$ is a very deep
recurrence, which would lead to poor time and space complexity. We
therefore need, as is done in the standard
Lanczos algorithm, to show that the recurrence equation can be
simplified.

We first restate the recurrence relations from the previous section, and
introduce some auxiliary notation $\bfc_{i+1,j}$.
\begin{align}
    \label{eq:blocklanczos:recurrence}
    \bfv_{i+1} &= A\bfv_i\bfd_i + \bfv_i(1-\bfd_i) - \sum_{j\leq i}
    \bfw_j
    \bfc_{i+1,j},\\
    \notag
    \bfc_{i+1,j}&=
    \bfw_j^{\text{inv}} \bfw_j^TA(A\bfv_i\bfd_i + \bfv_i(1-\bfd_i)).
\end{align}
Condition~\eqref{eq:blocklanczos:condition-orth} implies that the
second summand in the expression of $\bfc_{i+1,j}$ is zero for $j<i$.
We now examine
the first summand for $j<i$.
Consider equation~\eqref{eq:blocklanczos:recurrence} for index $j$, and
multiply by $\bfd_j$. We obtain:
\begin{align*}
    A\bfw_j&=\bfv_{j+1}\bfd_j+\cO(\langle\bfw_0,\ldots,\bfw_j\rangle),\\
    \bfw_j^TA^2\bfw_i &=\bfd_j\bfv_{j+1}^TA\bfv_i\bfd_i,
\end{align*}
where the notation $\cO(V)$, for $V$ a
subspace of $\F_p^N$, denotes any vector
block whose columns belong $V$.

The equations above yield the following simpler form for
$\bfc_{i+1,i-1}$:
$$\bfc_{i+1,i-1} 
    =\bfw_{i-1}^{\text{inv}} \bfv_i^TA\bfv_i\bfd_i.$$
Better, for $j+1<i$, we
consider
equation~\eqref{eq:blocklanczos:recurrence} for index $j+1$ and
multiply it by $1-\bfd_{j+1}$. We obtain the following equation, from
which we rewrite $\bfv_{j+1}$ in an interesting way:
\begin{align*}
    \bfv_{j+2}(1-\bfd_{j+1})&=\bfv_{j+1}(1-\bfd_{j+1}) +
    \cO(\langle\bfw_0,\ldots,\bfw_{j+1}\rangle),\\
    \bfv_{j+1} &= \bfw_{j+1} + \bfv_{j+1}(1-\bfd_{j+1})\\
    &= \bfv_{j+2}(1-\bfd_{j+1})  +
    \cO(\langle\bfw_0,\ldots,\bfw_{j+1}\rangle).
\end{align*}
This implies, for $j=i-1$, $j=i-2$, and more generally for any $j<i$
(repeatedly using the last fact):
\begin{align}
    \notag \bfc_{i+1,i-1} 
    &=\bfw_{i-1}^{\text{inv}} \bfv_i^TA\bfv_i\bfd_i,\\
    \notag \bfc_{i+1,i-2} &=\bfw_{i-2}^{\text{inv}}
(1-\bfd_{i-1})\bfv_{i}^TA\bfv_i\bfd_i,\\
\label{eq:blocklanczos:ci-general}
\bfc_{i+1,j} &= \bfw_j^{\text{inv}}
    \left(\prod_{k=j+1}^{i-1}(1-\bfd_{k})\right)\bfv_i^TA\bfv_i\bfd_i
\end{align}

We remark that this expression for $\bfc_{i+1,i-2}$ is simpler than 
in~\cite{EC:Montgomery95}.

In the normal course of the computation, it is easy to ensure that
$(1-\bfd_i)(1-\bfd_{i+1})=0$: this expresses the fact that column indices
which are not selected among the independent columns in $\bfv_i^TA\bfv_i$
used to define $\bfd_i$ have to be given priority when defining
$\bfd_{i+1}$ at the next step. As long as this can be achieved, we obtain
that whenever $j\leq i-3$, we have $\bfc_{i+1,j}=0$.
In~\cite{EC:Montgomery95}, Montgomery does exactly like this, and computes
each iterate $\bfv_{i+1}$ with access to only the
three previous iterates $\bfv_i$, $\bfv_{i-1}$, and $\bfv_{i-2}$.

The particular form of equation~\eqref{eq:blocklanczos:ci-general},
however, allows to write a simpler recurrence, which has the advantage of
limiting the storage needs of the algorithm. Define
\begin{align*}
    \bfp_i&=\bfv_{i-1}\bfw_{i-1}^{\text{inv}}
    +\bfv_{i-2}\bfw_{i-2}^{\text{inv}}(1-\bfd_{i-1})
    +\cdots,\\
    &=\sum_{j<i}\bfv_{j}\bfw_{j}^{\text{inv}}\prod_{k=j+1}^{i-1}(1-\bfd_k).
\end{align*}
By equations~\eqref{eq:blocklanczos:recurrence}
and~\eqref{eq:blocklanczos:ci-general}, we see that the contribution of
all iterates before $\bfw_i$ in the expression of $\bfv_{i+1}$ can be
simplified as:
\begin{equation*}
\sum_{j<i}\bfw_j\bfc_{i+1,j} = \bfp_i\bfv_i^TA\bfv_i\bfd_i.
\end{equation*}

The computation of the sequence of vector blocks
$(\bfv_i)_{i\geq0}$ from a starting vector block $\bfv_0$ can now be
summarized. At the start, we have $\bfp_0=0$.
For all $i\geq 0$, we proceed through the following steps.
\begin{itemize}
    \item Compute $\bfv_i^TA\bfv_i$ and
        $\bfv_i^TA^2\bfv_i$.
        Deduce $\bfd_i$ (giving, or not, priority to indices not selected
        in $\bfd_{i-1}$ --- it makes no difference) and $\bfw_i^{\text{inv}}$.
        If $\bfd_i=0$, terminate (see §\ref{sec:blocklanczos:termination}).
    \item Compute
        \begin{align*}
            \bfc_{i+1,i}&=\bfw_i^{\text{inv}} (\bfv_i^TA^2\bfv_i\bfd_i +
            \bfv_i^TA\bfv_i(1-\bfd_i)),\\
            \bfv_{i+1}&=A\bfv_i\bfd_i + \bfv_i(1-\bfd_i)
            -\bfv_i\bfc_{i+1,i}
            -\bfp_i\bfv_i^TA\bfv_i\bfd_i,\\
            \bfp_{i+1}&=\bfv_{i}\bfw_{i}^{\text{inv}}+\bfp_{i}(1-\bfd_{i})
        \end{align*}
    \item Memorize $\bfv_{i+1}$ and $\bfp_{i+1}$ for the next iteration.
\end{itemize}

\section{Termination}
\label{sec:blocklanczos:termination}

As the computation of the sequence of vector block proceeds, we clearly
have
\begin{align*}
    \langle\bfw_0,\ldots,\bfw_i\rangle&\subset\langle\bfv_0,A\bfv_0,\ldots,A^k\bfv_0,\ldots\rangle,\\
    \dim\langle\bfw_0,\ldots,\bfw_i\rangle&=\sum_{j\leq
i}\rank\bfd_j=\sum_{j\leq i}\rank\bfv_j^TA\bfv_j,\\
    \dim\langle\bfv_0,A\bfv_0,\ldots,A^k\bfv_0,\ldots\rangle&\leq N.
\end{align*}
Therefore, the number of iterations can be studied by first examining the
expected rank of $\bfv_j^TA\bfv_j$. Montgomery writes
in~\cite{EC:Montgomery95} the generating
function for the rank defect of an arbitrary $n\times n$ symmetric matrix
over $\F_p$. For $p=2$, the result obtained is that the expected rank
defect is $0.764...$. We thus have
$\Expect[\rank\bfd_i]\approx N-0.764$, from which we expect
that at most an expected value of $\frac{N}{n-0.764}$ iterations are
computed.

The actual termination condition which causes the iterative process to stop at index
$m$ (more
exactly, become stationary, if we consider $\bfw_{m}^{\text{inv}}=0$ to
be a legitimate value) is when we reach $\bfd_m=0$, which means that
$\bfv_m^TA\bfv_m=0$. When the block dimension $n$ is exceptionally small,
this might happen sooner than the expected value computed above, out of bad luck. We consider here that the block
dimension is large enough, so that this situation does not happen.

Heuristically, we expect a large intersection of
$\langle\bfv_m\rangle$ with the null space of $A$, which allows to find
close to $n$ solutions to the homogeneous linear system $Ax=0$.

We provide here some justification for this fact. Let $\bfdelta_0$ be a
vector block with $A\bfdelta_0=0$, and let $\bfv_0$ be an arbitrary
vector block.  We consider the two sequences corresponding to $\bfv_0$
and $\bfv'_0=\bfv_0+\bfdelta_0$.  It is easy to see that both sequences
evolve synchronously, as the matrices $\bfv_i^TA\bfv_i$ are equal for
both sequences at each step.  Let
$\bfDelta_i=\barvector{\bfv'_i-\bfv_i}{\bfp'_i-\bfp_i}$.
We have
\begin{align*}
\bfDelta_{i+1}&=
\bfDelta_{i}
\times\mathfrak{S}_i,\\
\mathfrak{S}_i &=
\barmatrix
{(1-\bfd_i)-\bfc_{i+1,i}}
    {\bfw_{i}^{\text{inv}}}
    {-\bfv_i^TA\bfv_i\bfd_i}
    {(1-\bfd_{i})}.
\end{align*}
We claim that $\mathfrak{S}_i$ is invertible. Indeed, we have:
\begin{gather*}
    \barmatrix{1-\bfd_i}{\bfd_i}{\bfd_i}{1-\bfd_i}\times
\mathfrak{S}_i \times
\barmatrix{1}{0}{\bfv_i^TA^2\bfv_i\bfd_i}{1-\bfd_i+\bfd_i\bfv_i^TA\bfv_i\bfd_i}
=\\
\barmatrix
{(1-\bfd_i)-
\bfd_i\bfv_i^TA\bfv_i\bfd_i}
    {0}
{-\bfw_i^{\text{inv}} (\bfv_i^TA\bfv_i(1-\bfd_i))}
{1}.
\end{gather*}
The latter matrix is clearly of full rank. A consequence is that
$\rank\bfDelta_m=\rank\bfDelta_0$, and that $\rank(\bfv'_m-\bfv_m)$ is
expected to be close to $\rank(\delta_0)$. We thus have no reason to expect
that $\bfv_m$ is an abnormally poor supply of elements of the null space
of $A$.

In the case which is relevant for integer factorization problems, we want
to solve the equation $x^TM=0$, and use the block Lanczos algorithm with
$A=MM^T$. In this case, we expect (as above, heuristically) that the intersection of
$\langle\bfv_i\rangle$ with the (left) null space of $M$ is large enough
to obtain close to  $n$ solutions to the linear system (provided the null
space itself is large enough).
\medskip

The block Lanczos algorithm can also be used to solve
inhomogeneous linear systems, to some extent.
In this case, we assume that the null space dimension is small compared
to the block dimension $n$. We want to solve $Ax=b$ for several vectors $b$.
We set the starting vector block $\bfv_0$ with our vectors $b$, and
complete with random vectors so as to form an $n$-dimensional vector block. We compute
$$\bfx=\sum_{i<m}\bfv_i\bfw_i^{\text{inv}}\bfv_i^T\bfv_0.$$
Note that $\bfx$ can be computed online at little extra cost, since for
$i\geq1$ we have
\begin{align*}
    \barvector{\bfv_0^T\bfv_{i+1}}{\bfv_0^T\bfp_{i+1}} &=
    \barvector{\bfv_0^T\bfv_{i}}{\bfv_0^T\bfp_{i}} \times\mathfrak{S}_i
\end{align*}
with $\mathfrak{S}_i$ as above.
Maintaining the evolution of $\bfx$ throughout the computation of the
sequence costs some extra memory.

By construction, we have $A\bfx-\bfv_0\in\langle\bfv_m\rangle$.
In~\cite{EC:Montgomery95}, Montgomery argues that heuristically, we have
$\langle A\bfv_m\rangle\subset\langle\bfv_m\rangle$. Based on the
assumption that the null space of $A$ is small enough, we hope to find
linear combinations of the columns of $A\bfx-\bfv_0$ and $A\bfv_m$ which
provide some solutions to $Ax=b$.

\section{Implementation in parallel}

Several implementations of the block Lanczos algorithm exist, and adapt
reasonably well to parallel computing environments. Different
processors (which can be different nodes communicating via message
passing, or simply processor threads) can collectively compute the
sequence $(\bfv_i)_{i\geq0}$.  It is useful to organize processors in a
two-dimensional (possibly toroidal) mesh, following the explanation in~\cite{Montgomery00}.
For simplicity, we assume the mesh has size $d\times d$.  Each processor
``owns'' part of the data: all vectors considered in the algorithm are
divided in $d^2$ fragments, and the matrix $M$ itself is also spread
across processes, in $d^2$ fragments.
An example organization, assuming that $M$ has dimension $N_1\times N_2$
(both assumed to be divisible by $d^2$), distributes data as follows for
the processor on row $i$ and column $j$ (both indexed from $0$) within the mesh:
\begin{itemize}
    \item For vector blocks of size $N_1\times n$, row indices $[x,x+\frac{N_1}{d^2}-1]$ with $x=(di+j)\frac{N_1}{d^2}$.
    \item For vector blocks of size $N_2\times n$, row indices $[x,x+\frac{N_2}{d^2}-1]$ with $x=(dj+i)\frac{N_2}{d^2}$.
    \item Sub-block of $M$ at position $(i,j)$ when split in blocks of
size $\frac{N_1}d\times\frac{N_2}d$.
\end{itemize}
In fact, load balancing has to be taken into account, so that the
distribution may actually be slightly different, or equivalently we may
need to permute rows and columns of $B$ adequately.

In this setting, many operations on vectors can be performed locally.  The
only collective operation at each step is the multiplication by $A=MM^T$,
decomposed into $u^T \leftarrow v^T M$ first, then $v \leftarrow M u$, where
$u$ and $v$ are vector blocks of size $N_2\times n$ and $N_1\times n$,
respectively. Communication goes as follows. After $u^T \leftarrow v^T M$,
processors on the same mesh column need to share their results so as to
form $N_2/d$ valid coefficients of the resulting vector $u$. For the
operation $v \leftarrow M u$, the processors in this same mesh column all
need these same $N_2/d$ input coefficients. Therefore, the communication 
operation required after each of these two products is in fact a pretty
common pattern. In the Message Passing Interface, this operation is called
``All-reduce'', and is usually well optimized and tuned on most serious
MPI implementations.

The other operations within each iteration are either of moderate cost (dot
products, or multiplication of vectors by $n\times n$ matrices) or totally
negligible (arithmetic directly involving $n\times n$ matrices). It should
be noted however that the parallelization of the block Lanczos algorithm
can only go as far as the communication speed allows, since synchronization
has to occur after each multiplication by $A=MM^T$.

\section{Recent developments}
\label{sec:blocklanczos:recent}

The block Lanczos algorithm has been successful in factoring projects since
its inception, including record computations until 2005. Compared to the
block Wiedemann algorithm~\cite{Coppersmith94}, the block Lanczos algorithm
seems to need a smaller number of multiplications of matrices by blocks of vectors.
With blocking dimension $n$, block Lanczos requires $2N/(n-0.764)$ products
in total (counting two for each iteration). The block Wiedemann algorithm
requires instead $\frac Nm+2\frac Nn$ products, depending on the two
blocking dimensions $m$ and $n$. When these are
chosen
straightforwardly as $m=n$, the algorithm needs $3N/n$ products. This
comparison can shift towards
being in favour of the block Wiedemann algorithm in two ways. First,
if for example when $m=4n$ is a valid choice, only $2.25N/n$
products are needed. Also, if large blocking dimensions can be considered
(say we use blocking dimensions $m'$ and $n'$ that are two appropriate multiples of $n$), then by
\cite[Theorem 7]{Kaltofen95}, only $(1+o(1))N/n$ products are needed,
which is better than block Lanczos.
However, the reason why the block Wiedemann algorithm has been
preferred in most factoring records since 2005 is simply because of the better
distribution opportunities it offers, a criterion which has been most
important given the composite nature of the hardware platforms used.

One may wonder whether the block Lanczos algorithm can be profitably used
in the context of the computation of discrete logarithms, in particular
with the number field sieve variants. An artifact of the number field sieve
for discrete logarithms, called Schirokauer maps, divides the presentation
of the linear algebra problem in two different settings. Given a sparse
matrix $M$ defined over a large finite field, the Schirokauer maps form a
dense matrix block $\bfS$ (with very few columns, but with large coefficients)
such that the linear system to be solved can be written as
$(M\mid\bfS)x=0$. It is not, however, the only way to proceed: any
vector $x$ such that $Mx\in\langle\bfS\rangle$ is a satisfactory solution.
As it turns out, this approach is viable both in the block Lanczos
algorithm, as discussed in §\ref{sec:blocklanczos:termination}, as well as in the block
Wiedemann algorithm, as discussed in~\cite[§8]{Coppersmith94}. In both
cases, this is possible as long as the number of columns of the block
$\bfS$ is less than the block dimension $n$.


\bibliographystyle{splncs03}
\bibliography{abbrev2,crypto_crossref,bib}

\begin{thebibliography}{10}
\providecommand{\url}[1]{\texttt{#1}}
\providecommand{\urlprefix}{URL }

\bibitem{BoLe16}
Bos, J.W., Lenstra, A.K.: Topics in Computational Number Theory inspired by
  Peter L. Montgomery. Cambdridge University Press (2016), to appear

\bibitem{Coppersmith93a}
Coppersmith, D.: Solving linear equations over $\mathrm{GF}(2)$: Block
  {Lanczos} algorithm. Linear Algebra Appl.  192,  33–60 (Jan 1993)

\bibitem{Coppersmith94}
Coppersmith, D.: Solving linear equations over $\mathrm{GF}(2)$ via block
  {Wiedemann} algorithm. Math. Comp.  62(205),  333–350 (Jan 1994)

\bibitem{EbKa97}
Eberly, W., Kaltofen, E.: On randomized {Lanczos} algorithm. In: Küchlin, W.W.
  (ed.) ISSAC~'97. p. 176–183. ACM Press (1997), extended abstract

\bibitem{Kaltofen95}
Kaltofen, E.: Analysis of {Coppersmith}'s block {Wiedemann} algorithm for the
  parallel solution of sparse linear systems. Math. Comp.  64(210),  777–806
  (Apr 1995)

\bibitem{C:LaMOdl90a}
LaMacchia, B.A., Odlyzko, A.M.: Solving large sparse linear systems over finite
  fields. In: Menezes, A.J., Vanstone, S.A. (eds.) CRYPTO'90. {LNCS}, vol. 537,
  pp. 109--133. Springer, Berlin, Germany, Santa Barbara, CA, USA (Aug~11--15,
  1990)

\bibitem{EC:Montgomery95}
Montgomery, P.L.: A block {Lanczos} algorithm for finding dependencies over
  gf(2). In: Guillou, L.C., Quisquater, J.J. (eds.) EUROCRYPT'95. {LNCS}, vol.
  921, pp. 106--120. Springer, Berlin, Germany, Saint-Malo, France (May~21--25,
  1995)

\bibitem{Montgomery00}
Montgomery, P.L.: Parallel block {Lanczos} (2000), slides of presentation at
  RSA-2000, dated January 17, 2000

\bibitem{PaTaLi85}
Parlett, B.N., Taylor, D.R., Liu, Z.A.: A look-ahead {Lanczos} algorithm for
  unsymmetric matrices. Math. Comp.  44(169),  105–124 (Jan 1985)

\bibitem{Wiedemann86}
Wiedemann, D.H.: Solving sparse linear equations over finite fields. IEEE
  Trans. Inform. Theory  IT–32(1),  54–62 (Jan 1986)

\end{thebibliography}

\end{document}